# Discovery of Superconductivity in (Ba,K)SbO$_3$

Minu Kim[1], Graham M. McNally[1], Hun-Ho Kim[1], Mohamed Oudah[2], Alexandra Gibbs[3], Pascal Manuel[3], Robert Green[2,4], Tomohiro Takayama[1], Alexander Yaresko[1], Ulrich Wedig[1], Masahiko Isobe[1], Reinhard K. Kremer[1], D. A. Bonn[2], Bernhard Keimer[1], and Hidenori Takagi[1,5]

[1]Max Planck Institute for Solid State Research, Heisenbergstrasse 1, 70569 Stuttgart, Germany.

[2]Stewart Blusson Quantum Matter Institute, University of British Columbia, Vancouver, British Columbia V6T 1Z4, Canada.

[3]ISIS Facility, STFC Rutherford Appleton Laboratory, Harwell Science and Innovation Campus, Oxon OX11 0QX, United Kingdom.

[4]Department of Physics & Engineering Physics, University of Saskatchewan, Saskatoon, Saskatchewan S7N 5E2, Canada.

[5]Department of Physics, University of Tokyo, Bunkyo-ku, Hongo 7-3-1, Tokyo 113-0033, Japan.

Superconducting bismuthates (Ba,K)BiO$_3$ (BKBO) constitute an interesting class of superconductors in that superconductivity with a remarkably high $T_c$ of 30 K[1,2] arises in proximity to charge density wave (CDW) order. Prior understanding on the driving mechanism of the CDW and superconductivity emphasizes the role of either bismuth (negative $U$ model)[3,4] or oxygen ions (ligand hole model)[5,6]. While holes in BKBO presumably reside on oxygen owing to their negative charge transfer energy, so far there has been no other comparative material studied. Here, we introduce (Ba,K)SbO$_3$ (BKSO) in which the Sb 5$s$ orbital energy is higher than that of the Bi 6$s$ orbitals enabling tuning of the charge transfer energy from negative to slightly positive. The parent compound BaSbO$_{3-\delta}$ shows a larger CDW gap compared to the undoped bismuthate BaBiO$_3$. As the CDW order is suppressed via potassium substitution up to 65 %, superconductivity emerges, rising up to $T_c$ = 15 K. This value is lower than the maximum $T_c$ of BKBO, but higher by more than a factor of two at comparable potassium concentrations. The discovery of an enhanced CDW gap and superconductivity in BKSO indicates that the sign of the charge transfer energy may not be crucial, but instead strong metal-oxygen covalency plays the essential role in constituting a CDW and high-$T_c$ superconductivity in the main-group perovskite oxides.

Superconducting bismuthates, BaPb$_{1-x}$Bi$_x$O$_3$ (BPBO)[7] and Ba$_{1-x}$K$_x$BiO$_3$ (BKBO)[1,2], have attracted considerable research interest since their discovery more than three decades ago. The parent compound BaBiO$_3$ (BBO) is known to be a non-magnetic, commensurate charge density wave (CDW) insulator. The CDW order is accompanied by a breathing octahedral distortion, that is, two octahedra with different sizes order in a three-dimensional checkerboard pattern[8,9]. As the CDW order is suppressed via chemical substitution of Ba with K or Bi with Pb[10,11], the compounds become superconducting up to a maximum $T_c$ of 12 K in BPBO, and 30 K in BKBO. Numerous experiments have established that the mechanism of superconductivity is largely conventional; the pairing symmetry is $s$-wave[12], and the oxygen isotope effect is consistent with the BCS theory[13], meaning electron-phonon interaction plays the important role in superconductivity. Nevertheless, the unexpectedly high $T_c$ of BKBO, despite a rather low carrier density, has triggered various critical questions as to the driving mechanism and the correct model of the CDW order and superconductivity in these materials. Recent studies suggest that the additional consideration of long-range exchange interactions and many-body effects can be crucial for the quantitative description of the CDW gap[14,15] as well as for superconductivity[16-18].

Historically, superconducting bismuthates have been considered as archetypal candidates for unconventional superconductors in which an effective attractive electron-electron interaction (negative $U$[19,20]) leads to electron pairing in real as well as in momentum spaces[3,4]. The real-space pairing occurs in the parent compound BBO, which is argued to be a typical valence-skipping compound with unstable tetravalent bismuth ($6s^1$) disproportionated into tri- ($6s^2$) and pentavalent ($6s^0$). As the charge disproportionation on the bismuth sites is suppressed via chemical doping and the bismuth valence starts to dynamically fluctuate, negative $U$, which causes pairing of two electrons in the Bi$^{3+}$-O$_6$ octahedra, could also pair them in $k$-space, leading to bipolaronic superconductivity. Although its direct evidence still remains to be clarified, the model has provided a possible framework to understand superconductivity in the bismuthates as well as some chalcogenides[21] with valence-skipping elements.

While the negative $U$ model emphasizes the role of bismuth, an alternative model asserts the role of oxygen and its hybridization with bismuth[5,6,10,22]. Its foundation is that the charge transfer energy $\Delta_{CT}$ of the bismuthates is negative, as the on-site energy of the Bi 6$s$ orbital is lower than that of the oxygen 2$p$ owing to the large relativistic effect of heavy bismuth[23]. Consequently, electronic states around the Fermi level (which originate from the strongly hybridized $sp\sigma^*$ antibonding states) show predominantly oxygen 2$p$ character. This crucially modifies the preceding understanding of the CDW order in BBO; it should be described not by the charge disproportionation ($6s^2 + 6s^0$) but rather by the bond-length disproportionation as $6s^2 + 6s^2\underline{L}^2$, where $\underline{L}$ denotes a ligand hole. Spectroscopic evidence supports the oxygen hole model[24,25]. As the CDW order is suppressed, oxygen holes become delocalized, giving rise to superconductivity, possibly via strong electron-phonon coupling[6,26]. The importance of oxygen holes has previously been demonstrated in the Zhang-Rice model[27] for cuprates, in which Cu $d$ electrons and oxygen holes form a strongly hybridized singlet state, highlighting their potential role in understanding the CDW order and high-$T_c$ superconductivity in the bismuthates as well.

In spite of their scientific importance, a contrastive analysis of the effects of bismuth and oxygen has so far been limited due to lack of compounds analogous to the bismuthates. Perovskite antimonates are ideal candidates to study, because antimony is isovalent to bismuth. In addition, higher on-site energy of the Sb 5$s$ orbital compared to the Bi 6$s$ may enable us to tune $\Delta_{CT}$ of the material from negative to positive, therefore orbital characters of the states around the Fermi level are expected to be modified. Several attempts have been made to synthesize superconducting antimonates, but with limited success. Cava et al. reported that partially doped $BaPb_{0.75}Sb_{0.25}O_3$ becomes superconducting[28], but its $T_c$ is significantly decreased as compared to $BaPb_{0.75}Bi_{0.25}O_3$. The role of antimony has also been investigated via first-principles calculation, revealing a possible strong electron-phonon coupling in antimonates[29]. However, superconducting perovskite antimonates, with only antimony occupying the octahedral sites of perovskites, have yet to be experimentally reported. This is probably because the strongly covalent Sb-O bond is known to hamper forming 180 degree Sb-O-Sb bonds[30], and as a consequence, no perovskite antimonates have been realized up to date except a highly distorted insulating $NaSbO_3$[31]. Here, we report new superconducting antimonates $Ba_{1-x}K_xSbO_3$ (BKSO), which we were able to stabilize for the first time via high-pressure high-temperature synthesis routes, enabling clarification of possible driving mechanisms for CDW and superconductivity in the compounds by comparing their properties with the sibling compound, BKBO.

In order to shed light on the effects that varying $\Delta_{CT}$ has on the electronic structure, the band structures of BKSO as well as BKBO are calculated via the hybrid-DFT method[14], as shown in Fig. 1. Here we used their atomic structures without the breathing distortions, which were experimentally obtained at the potassium concentration $x$ of ~0.65 as discussed later. BKBO is anticipated to have negative $\Delta_{CT}$[5,6], as explained earlier. Therefore, it should show predominant oxygen 2p character in the $sp\sigma^*$ antibonding band, and Bi 6s character in the $sp\sigma$ bonding band. The distinct characters of the two bands are clearly confirmed in the band structure calculation (For more quantitative analysis, see Extended Data Fig. 1). In contrast, we find that BKSO shows stronger Sb 5s character in the $sp\sigma^*$ band as compared to BKBO. This suggests $\Delta_{CT}$ of BKSO may be marginally positive, which is further supported by Wannierization analysis (Extended Data Fig. 2 and Table 1). In addition, both Sb 5s and oxygen 2p characters are rather mixed in the bonding and antibonding bands (also see Extended Data Fig. 1), corroborating stronger metal-oxygen covalency in BKSO.

Having clarified the inverted $\Delta_{CT}$ of BKSO as compared to that of BKBO, we next reveal how the inversion modifies physical properties of the materials. First, the parent compound BaSbO$_{3-\delta}$ (BSO) is found to be a robust insulator with a larger CDW gap. Rietveld refinement of neutron powder diffraction data (Extended Data Fig. 3a) confirms cubic symmetry (space group $Fm\bar{3}m$) with the breathing distortion, and furthermore, reveals two distinct Sb-O bond lengths, 2.24(1) and 2.01(1) Å (Fig. 2). Surprisingly, its bond-length difference ($\Delta d$ = 0.23 Å) is larger than that of BBO ($\Delta d$ = 2.28 − 2.11 = 0.16 Å)[8]. The observation of the breathing distortion establishes the commensurate CDW order from the structural point of view, which results in a band gap in the material. Using diffusive reflectance spectroscopy, the band gap in BSO is determined to be 2.54 eV. The value is significantly larger than that in BBO (2.02 eV)[32], consistent with the larger bond-length disproportionation observed in the neutron diffraction.

The CDW order of the antimonate can be manageably suppressed by substituting Ba with potassium up to 65 %. X-ray and neutron powder diffraction measurements enable us to map out the structural phase diagram of BKSO, as shown in Fig. 3 (For detailed refinement profiles, see Extended Data Figs. 3 and 4). As $x$ increases, two structural transitions are found, namely, from $Fm\bar{3}m$ to $I4/mcm$ at $x \approx 0.3$, and from $I4/mcm$ to a $Pm\bar{3}m$ phase at $x \approx 0.65$. Apparently, this structural phase diagram is qualitatively similar to that of BKBO[33], in which the CDW order is sequentially suppressed from the long-range to short-range and then finally disappears. The structural transitions also trigger a drastic change in Raman scattering, as shown in Fig. 3b. The undoped compound shows a pronounced peak at 672 cm$^{-1}$, which corresponds to the breathing-mode phonon, that is, the symmetric movement of oxygen ions with $A_{1g}$ symmetry and also known to be crucial in superconductivity in BKBO[34]. The phonon peak is first marginally softened as $x$ increases, and next its amplitude completely vanishes for $x \geq 0.65$. Because all the phonon modes become Raman-inactive in the $Pm\bar{3}m$ phase[35], the vanishing breathing-mode peak confirms that the crystal symmetry above $x = 0.65$ is indeed $Pm\bar{3}m$ and that the CDW order is completely suppressed. Interestingly, the critical potassium concentration $x_{IMT}$ at which the structure symmetry becomes $Pm\bar{3}m$ is larger in BKSO ($x_{IMT} \sim 0.65$) than in BKBO ($x_{IMT} \sim 0.35$). The larger $x_{IMT}$ is plausibly related to the bigger CDW gap, which may necessitate more holes for its suppression in the antimonates.

BKSO, with $x = 0.67$ and $Pm\bar{3}m$ symmetry, shows bulk superconductivity with a maximum $T_c$ of 15.0 K, as depicted in Fig. 4. Specific heat of the optimally doped sample ($x = 0.67$) shows a jump at 15.0 K due to the superconducting transition, which is suppressed by applying a magnetic field of 1 T (Fig. 4a). While the jump is broadened, perhaps indicating sample inhomogeneity from the high-pressure synthesis, the electronic specific heat $\Delta C/T$ shows a change on the order of $\gamma$ (0.924 mJ·mol$^{-1}$K$^{-2}$; Extended Data Fig. 5), indicating bulk superconductivity. Further convincing evidence of bulk superconductivity can be found in magnetic susceptibility, with a diamagnetic volume fraction near 100 % (Fig. 4b). Importantly, the $T_c$ of BKSO is found to be lower than the maximum $T_c$ of BKBO (~30 K at $x = 0.4$), but at the same potassium concentration, the $T_c$ is higher than that of BKBO (7.0 K at $x = 0.66$)[36] by more than factor of two.

Reduced oxygen hole character in the optimally doped BKSO is revealed via X-ray absorption spectroscopy (XAS), consistent with its positive $\Delta_{CT}$. The oxygen $K$-edge, which probes unoccupied oxygen 2p states, is measured and compared with a BKBO reference at similar potassium concentration, as shown in Fig. 4c. As reported previously, BKBO exhibits a significant pre-peak at $E$ = 528.5 eV, which indicates predominant oxygen holes in the conduction band[37]. Interestingly, we find that BKSO shows the pre-peak at the same energy but its intensity is diminished. It should be mentioned that the Sb $M_5$-edge gives rise to an additional feature in the spectrum because of its similar energy to that of the oxygen $K$-edge[38], complicating further quantitative analysis at present.

A phase diagram of BKSO, compiled from these results, offers a comprehensive view of the interplay between the CDW order and superconductivity in main-group oxide superconductors (Fig. 5). First, a common tendency in the phase diagram can be found in both compounds; as the CDW insulating phase is suppressed, a half-dome of superconductivity arises with $T_c$ maximized at the border of the insulator-to-metal transition and gradually decreasing with $x$ increased. Nevertheless, a crucial difference between the two compounds is that the suppression of the CDW phase occurs at higher $x$ in BKSO, possibly related with its larger CDW gap. As BKSO shows higher $T_c$ at $x \geq 0.65$, its $T_c$ could have exceeded that of BKBO if it were possible to stabilize metallic BKSO at lower $x$ and the same trend of $T_c$ on $x$ held. In reality, this is so far prohibited by the strong CDW instability in BKSO, setting a limit on enhancing superconductivity further.

Though its $T_c$ is lower than the maximum value of BKBO, enhanced superconductivity in BKSO at $x$ = 0.65 suggests a mechanism associated with strong metal-oxygen covalency. Strong covalency in BKSO can be basically understood as a consequence of decreased absolute value of $\Delta_{CT}$, making the orbital energies of Sb 5$s$ and O 2$p$ very close (Fig. 1d). As a result, Sb-O bonds are expected to be stiffer than Bi-O bonds, giving rise to increased phonon energy, which is indeed supported by experiments: First, the Debye temperature of the optimally doped BKSO is found to be 535 K (Extended Data Fig. 4), which is greater than that of BKBO, typically ~330 K[39]. Second, the breathing-mode phonon in the parent compounds show its frequency increased by about 19 % in BSO (672 cm$^{-1}$), as compared to BBO[35] (565 cm$^{-1}$; Extended Data Fig. 6). Taking a simple approach assuming that superconductivity in BKSO arises from electrons strongly coupled to the high-energy breathing phonon, our results indicate that electron-phonon coupling may be stronger in BKSO than in BKBO (see Method section for more detailed discussion). Larger CDW instability in BKSO could also be related to its stronger electron-phonon coupling, which is reminiscent of covalent superconductors in which too large electron-phonon coupling sometimes leads to lattice instability instead of higher $T_c$[40,41].

Finally, yet importantly, our study affords a new insight into the role of negative $\Delta_{CT}$ and oxygen holes. On one hand, it is evident from the results that BKSO, which has positive $\Delta_{CT}$, demonstrates both CDW insulating and superconducting phases, analogous to BKBO. Therefore, the emergence of the two phases is not necessarily dependent on the sign of $\Delta_{CT}$. In addition, at $x \geq 0.65$ BKSO shows even higher $T_c$ than BKBO, in spite of its decreased oxygen hole character. Hence, it could be inferred that oxygen holes may not be a *necessary* condition for enhancement of superconductivity. On the other hand, BKBO, which has more oxygen holes, shows the unexpectedly smaller amplitude of the CDW order than that in BKSO, which results in a smaller $x_{IMT}$ as well. Since the electronic DOS as well as electron-phonon interaction become larger with decreasing $x$[18], it would be naturally anticipated that a smaller $x_{IMT}$ may lead to an increased $T_c$. Therefore, the weakened CDW order in BKBO, possibly by oxygen holes, could be vital to show a higher $T_c$. In this respect, oxygen holes may be a *sufficient* condition for a higher $T_c$[42].

In conclusion, we have reported the discovery of new superconducting perovskite antimonates with a maximum $T_c$ of 15 K. The modification of $\Delta_{CT}$ via the substitution of Bi to Sb has allowed us to address long-standing questions as to the different roles of metal and oxygen ions for the CDW and superconductivity in the main-group oxide superconductors. Furthermore, these results provide fascinating possibilities for approaching novel regimes in the future. It would be interesting to change $\Delta_{CT}$ either to be more positive or negative by choosing appropriate elements at the octahedral site: arsenic and tin would give an on-site energy of *ns* level higher than antimony, thus their $\Delta_{CT}$ would be more positive than antimonates. Assuming these compounds can be stabilized, they would show more prominent effects of the cations, providing ideal model systems to examine the negative *U* model and the effects of valence fluctuations[3]. On the other hand, tellurium[43] and iodine[44] would give an on-site energy of *ns* level lower than antimony, and perhaps comparable or even lower than bismuth. Thus their $\Delta_{CT}$ would be more negative than antimonates and possibly bismuthates. Depending on their energy level relative to the Bi 6*s* level, they could provide additional model systems to comprehensively examine the effect of oxygen holes[5].

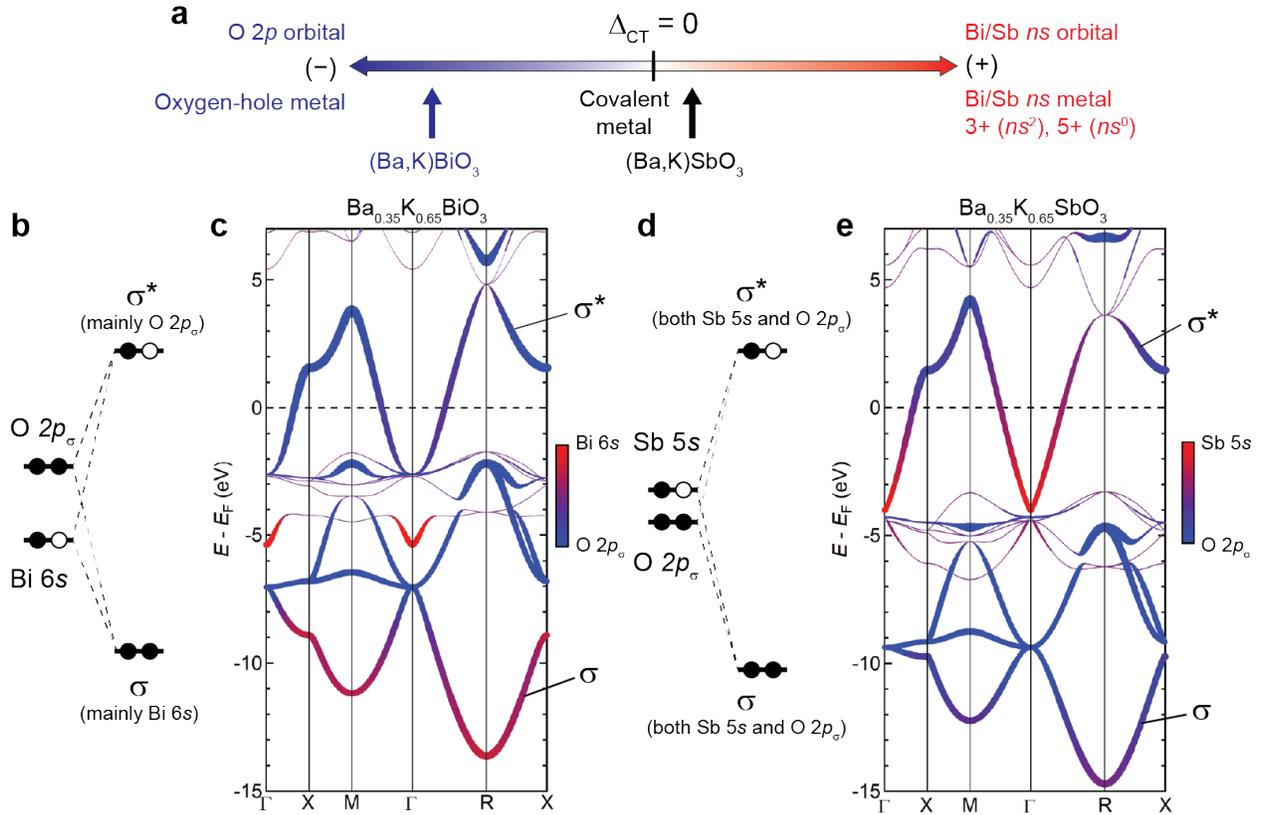

**Figure 1 | Inverted charge transfer energy of the antimonate as compared to the bismuthate. a,** A schematic diagram of different regimes of metallicity in Ba$_{1-x}$K$_x$BiO$_3$ (BKBO) and Ba$_{1-x}$K$_x$SbO$_3$ (BKSO). When charge transfer energy $\Delta_{CT}$ is positive, Bi or Sb $s$ electrons are dominant. On the contrary, when it is negative, oxygen holes become dominant. BKBO is presumably located in the scheme of the oxygen-hole metal ($\Delta_{CT}$ is negative)[5,6], because the Bi 6$s$ orbital energy is markedly lower than the O 2$p$ energy, as illustrated in its molecular orbital diagram, **b**. **c,** The fat-band representation of its electronic band structure calculated with hybrid-DFT confirms the predominant oxygen (bismuth) character in the σ* (σ) band, consistent with that expected from negative $\Delta_{CT}$. Meanwhile, BKSO is located rather in the scheme of the Bi/Sb $s$-orbital metal, in which the charge transfer energy is slightly positive while being close to the middle ($\Delta_{CT} \gtrsim 0$). This is due to the higher Sb 5$s$ orbital energy as depicted in its molecular

orbital diagram, **d**. The diagram is supported by the hybrid-DFT calculations, **e**. BKSO clearly shows a relatively stronger Sb 5$s$ character in the $\sigma^*$ band as compared to BKBO, while both Sb 5$s$ and oxygen 2$p$ are found to be highly mixed in both $\sigma$ and $\sigma^*$ bands, indicating strong Sb-O covalency.

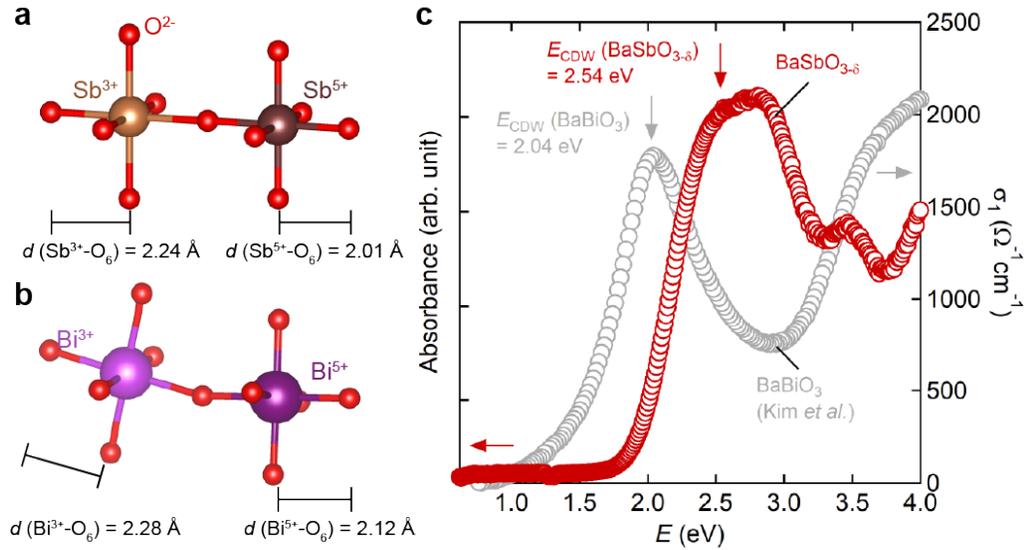

**Figure 2 | Three-dimensional charge density wave order in undoped BaSbO$_{3-\delta}$.** Schematic diagrams of expanded and contracted octahedra in **a,** BaSbO$_{3-\delta}$ and **b,** BaBiO$_3$. From the neutron diffraction investigations, two distinct Sb-O bond lengths are estimated to be 2.24(1) and 2.01(1) Å, respectively. Its bond-length difference is found to be larger than that of BaBiO$_3$[8]. **c,** Optical absorbance of BaSbO$_{3-\delta}$ at 300 K shows a wide band gap of 2.54 eV caused by the formation of the charge-density-wave (CDW) order. For the comparison, the optical conductivity of BaBiO$_3$[45] is plotted as a reference, indicating the CDW gap of BaSbO$_{3-\delta}$ is larger than that of BaBiO$_3$.

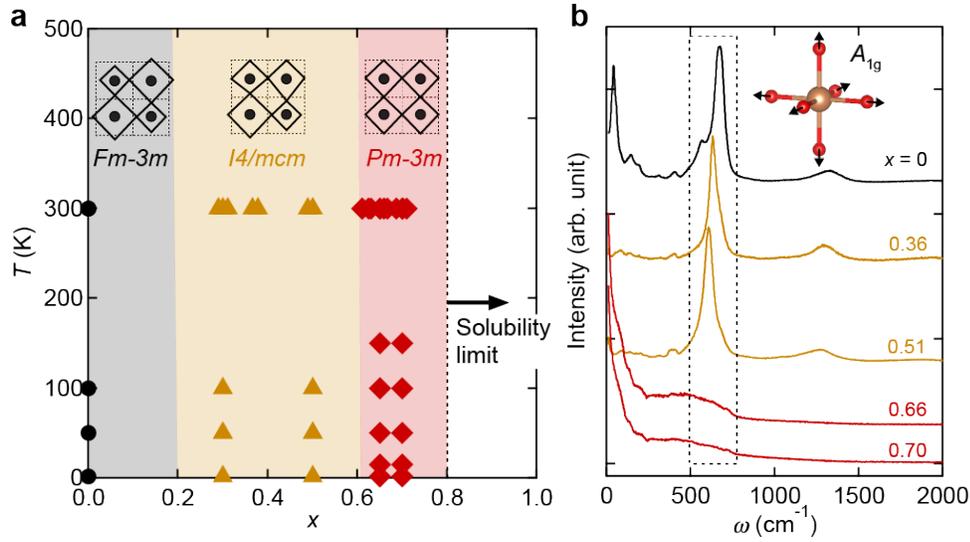

**Figure 3 | Suppression of the charge density wave order via potassium doping. a,** The structural phase diagram of $Ba_{1-x}K_xSbO_3$ based on X-ray and neutron diffraction data. Black dots, khaki triangles, and red diamonds represent the $Fm\bar{3}m$, $I4/mcm$, and $Pm\bar{3}m$ phases, respectively. The insets depict the local atomic structure of each phase, which shows transitions of the charge density wave (CDW) order from the commensurate long-range ($Fm\bar{3}m$) to short-range ($I4/mcm$), and finally to complete suppression ($Pm\bar{3}m$). **b,** Raman scattering, measured with the excitation wavelength of 632 nm at 300 K, shows that the breathing-mode phonon peak observed in $Fm\bar{3}m$ and $I4/mcm$ phases disappears in the $Pm\bar{3}m$ phase ($x \geq 0.65$), confirming the CDW order is completely suppressed. The inset shows a schematic picture of the breathing-mode phonon.

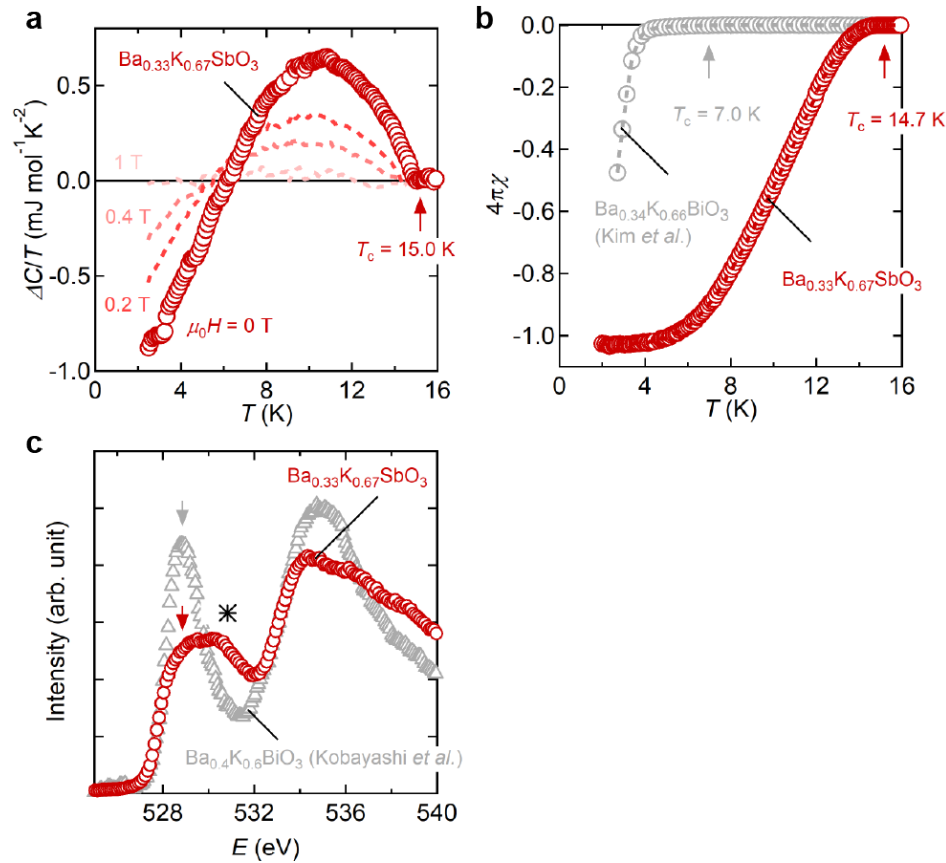

**Figure 4 | Bulk superconductivity and weakened oxygen hole character in the optimally doped $Ba_{1-x}K_xSbO_3$. a,** Superconducting transition observed in the specific heat of the optimally doped antimonate ($x = 0.67$). $\Delta C$ denotes the difference between specific heats under each field and 14 T. Superconducting transition temperature $T_c$ is estimated to 15.0 K from the jump, which can be suppressed by applying a field of 1 T. **b,** The superconducting transition of the same sample is observed in zero-field-cooled magnetic susceptibility at 0.001 T (red), in comparison with that of $Ba_{0.34}K_{0.66}BiO_3$ (grey)[36]. The diamagnetic volume fraction is near 100 %, indicating bulk superconductivity. Here, $T_c$ is defined as a temperature where the volume fraction started increasing by 0.1 %. **c,** Oxygen $K$-edge X-ray absorption spectrum of $Ba_{0.33}K_{0.67}SbO_3$ (red open circles) at 300 K, plotted together with that of $Ba_{0.4}K_{0.6}BiO_3$[37] (grey open triangles). The intensity of each spectrum was calibrated such that the intensity at a higher energy (~550 eV)

above the edge between two samples is to be equal. The decreased prepeak in the antimonate, shown by arrows, indicates the decrease of oxygen holes compared to the bismuthate. The peak indicated by the asterisk may be due to the Sb $M_5$-edge, whose energy is close to the oxygen $K$-edge.

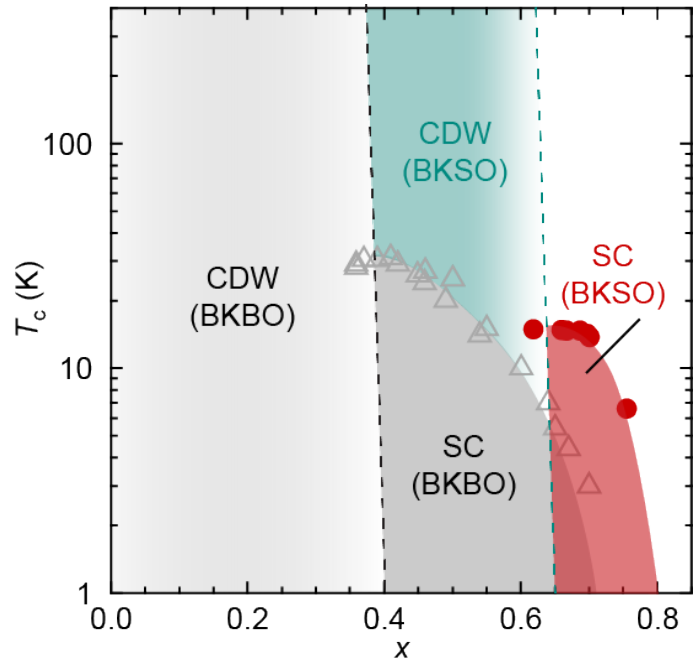

**Figure 5 | Phase diagram of charge density waves (CDW) and superconductivities (SC) in $Ba_{1-x}K_xSbO_3$ (BKSO) and $Ba_{1-x}K_xBiO_3$ (BKBO).** Red circles are the $T_c$'s of the antimonates with the same notation as in Figure 3, and grey triangles are $T_c$ of $Ba_{1-x}K_xBiO_3$[36,46]. $T_c$ of the antimonates show a half-dome shape (red region), which is similar to that of the bismuthates (dark grey). The crucial difference is that the CDW order in the bismuthates (light grey region) is suppressed at $x = 0.4$, whereas that of the antimonates (green region) continues to exist up to $x = 0.65$.

**Methods**

**Sample synthesis and characterization** (Ba,K)SbO$_3$ samples were fabricated using a high-pressure high-temperature synthesis technique with a Walker-type multi-anvil module. A typical synthesis condition was ~1200 ºC at 12 GPa. Powder X-ray diffraction was measured in the Debye-Scherrer geometry using a Mo $K\alpha_1$ source. Neutron powder diffraction was measured with a time-of-flight neutron source using the instrument WISH at ISIS. Optical absorbance was measured via diffuse reflectance spectroscopy at room temperature. Magnetic susceptibility and heat capacity were measured via a Quantum Design magnetic property measurement system (MPMS) and physical property measurement system (PPMS), respectively. X-ray absorption spectroscopy (XAS) is measured in partial fluorescence yield mode using a silicon drift detector to select the O $K$-edge fluorescence, at the Spherical Grating Monochromator (SGM) beamline at the Canadian Light Source.

**First-principles calculation** The band structures of BKSO and BKBO were calculated using the WIEN2k code[47] with full hybrid functionals (YS-PBE0, similar to HSE06[48]). For the calculations, 10×10×10 $k$ mesh, $R_{MT}K_{MAX}$ = 7.0, and the virtual crystal approximation were used. The atomic structures reported from experiments were used.

**Comparison of $T_c$ between BKSO and BKBO at $x$ = 0.65** Here we assume that superconductivity in both BKSO and BKBO is primarily caused by electrons strongly coupled with the high-frequency breathing phonon. According to the McMillan-Hopfield expression[49,50] for superconductors in strong electron-phonon coupling regime, $T_c$ is given as $\sqrt{\lambda\langle\omega^2\rangle} \sim \sqrt{\eta/M}$, where $\lambda$ is the electron-phonon coupling constant, $\langle\omega^2\rangle$ is the average of squared phonon frequency, $M$ is the atomic mass, and $\eta$ is the electronic spring constant. Also, $\lambda$ is given as

$N(E_F)\langle I^2\rangle/M\langle\omega^2\rangle$, where $N(E_F)$ is the DOS at the Fermi energy, $\langle I^2\rangle$ is the Fermi surface average of squared electron-phonon coupling interaction. For the cases of BKSO and BKBO, $M$ would be the mass of oxygen. Total DOS is estimated via the band calculation to be nearly identical in BKSO and BKBO (Extended Data Fig. 1). Then, the above expression of $T_c$ can be simplified to $T_c \propto \sqrt{\langle I^2\rangle}$. Therefore, increased $T_c$ in BKSO can be understood by increased $\langle I^2\rangle$.

**Acknowledgement** We thank P. Adler, K. Foyevtsova, G. Khaliullin, H. Mizoguchi, J.-G. Park, and J. Yu for discussions, and U. Engelhardt, F. Falkenberg, W. Kain, K. Schunke, and S. Strobel for experimental support. This research was carried out owing in part to funding from the Max Planck-UBC-UTokyo Centre for Quantum Materials. The Canadian Light Source (CLS) is funded by the Canada Foundation for Innovation, NSERC, the National Research Council of Canada, the Canadian Institutes of Health Research, the Government of Saskatchewan, Western Economic Diversification Canada, and the University of Saskatchewan. We thank the STFC ISIS facility for the provision of beamtime.

**Author Contributions** H.T. and M.K. conceived the project. M.K. prepared and characterized the samples. T.T., M.I., and R.K.K. helped the analysis. M.K., G.M.M., A.G., and P.M. conducted the neutron diffraction experiments. M.K. and H.-H.K. conducted the Raman experiments. M.K. and U.W. and A.Y. performed the first-principles calculations. M.O. and R.G. conducted the XAS experiments. M.K. and H.T. wrote manuscript and all authors commented on it.

**Competing interests** The authors declare no competing interests.

**Additional information**

**Correspondence and requests for materials** Should be addressed to M.K. or H.T.

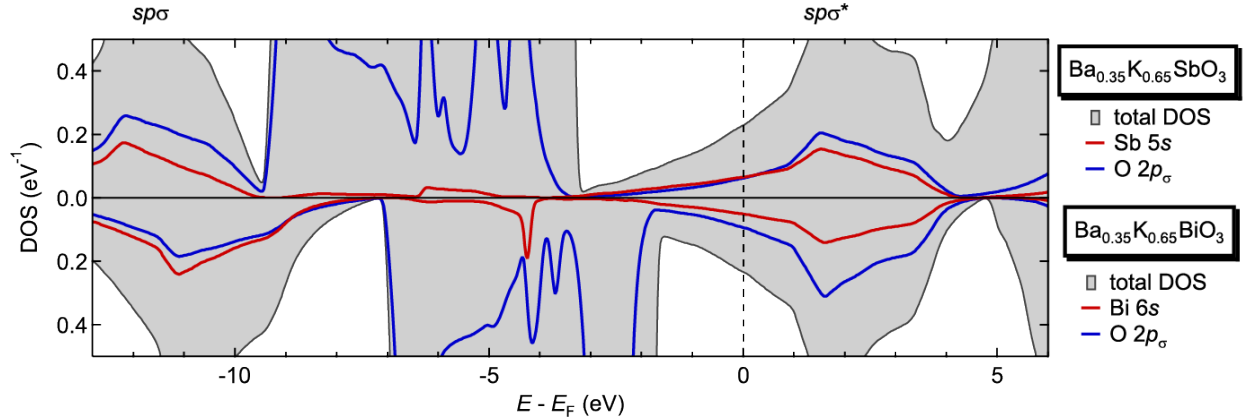

**Extended Data Figure 1 | Electronic density of states of BKSO and BKBO calculated via the hybrid-DFT method.** While total DOS at Fermi level shows almost no difference between the two compounds (BKSO: 0.229 states/eV, BKBO: 0.230 states/eV), projected DOS (PDOS) of metal $s$ and oxygen $2p_\sigma$ show meaningful differences: in BKBO, PDOS of O $2p_\sigma$ is bigger than Bi $6s$ [PDOS (Bi $6s$) / PDOS (O $2p_\sigma$) $\cong$ 0.548]. On the contrary, in BKSO, PDOS of Sb $5s$ and O $2p_\sigma$ at the Fermi level are almost equal in BKSO [PDOS(Sb $5s$) / PDOS (O $2p_\sigma$) $\cong$ 1.06]. The results are consistent with those suggested from the molecular-orbital diagram (Fig. 1); BKBO with negative $\Delta_{CT}$ shows predominant Bi $6s$ and O $2p_\sigma$ characters in the $sp\sigma$ bonding and $sp\sigma^*$ antibonding bands, respectively. On the contrary, BKSO, with $\Delta_{CT}$ slightly positive while close to zero, shows that the ratio between Sb $5s$ and O $2p_\sigma$ PDOS is largely unchanged in the bonding and antibonding bands. In addition, PDOS of the two orbitals at the Fermi level are almost equal.

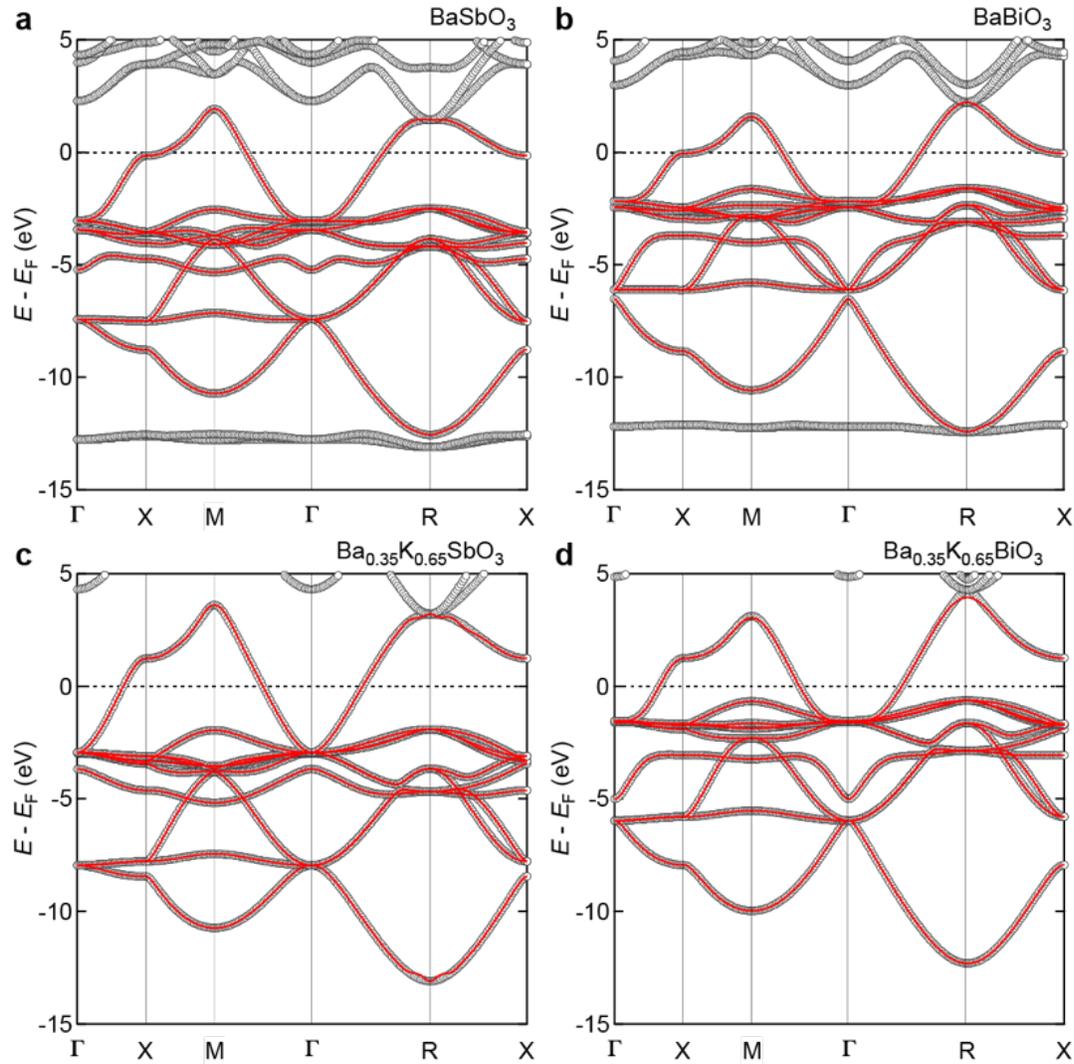

**Extended Data Figure 2 | Wannierization of BKSO and BKBO without and with potassium doping.** For Wannierization, density functional calculation on the band structures of **a**, BaSbO$_3$, **b**, BaBiO$_3$, **c**, Ba$_{0.35}$K$_{0.65}$SbO$_3$, **d**, Ba$_{0.35}$K$_{0.65}$BiO$_3$ were performed with the WIEN2k code using the PBE GGA (generalized gradient approximation) exchange correlation (black open circles). Other conditions are same as described in Methods. Wannierization using 7 x 7 x 7 $k$ mesh and 10 Wannier projections (1 × M $ns$ + 3 × oxygen 2$p_x$, 2$p_y$, 2$p_z$) was performed to fit the DFT results (red curves).

| Compound | $E_s$ (eV) | $E_{p\sigma}$ (eV) | $E_{p\pi}$ (eV) | $t_{sp\sigma}$ (eV) | $\Delta_{CT} = E_s - (E_{p\sigma} + 2E_{p\pi})/3$ (eV) |
|---|---|---|---|---|---|
| BaSbO$_3$ | −4.029 | −5.670 | −3.849 | 1.964 | 0.427 |
| BaBiO$_3$ | −5.298 | −4.612 | −2.796 | 1.995 | −1.897 |
| Ba$_{0.35}$K$_{0.65}$SbO$_3$ | −2.586 | −6.045 | −3.577 | 2.127 | 1.814 |
| Ba$_{0.35}$K$_{0.65}$BiO$_3$ | −3.634 | −4.213 | −2.017 | 2.150 | −0.885 |

**Extended Data Table 1 | Obtained parameters of BKSO and BKBO from Wannierization.** $E_s$, $E_{p\sigma}$, $E_{p\pi}$ are the onsite energies of M $ns$, O $2p_\sigma$, and O $2p_\pi$ orbitals, respectively. $t_{sp\sigma}$ is the hopping parameter between M $ns$ and O $2p_\sigma$, and $\Delta_{CT}$ is charge transfer energy defined as energy difference between the M $ns$ orbital and the average of O $2p_\sigma$ and O $2p_\pi$. BSO and BKSO show positive $\Delta_{CT}$, whereas BBO and BKBO show negative $\Delta_{CT}$. Increased $\Delta_{CT}$ in the doped compounds as compared to the undoped ones could be related to the effect of larger hybridization, as indicated by larger $t_{sp\sigma}$ in the doped compounds.

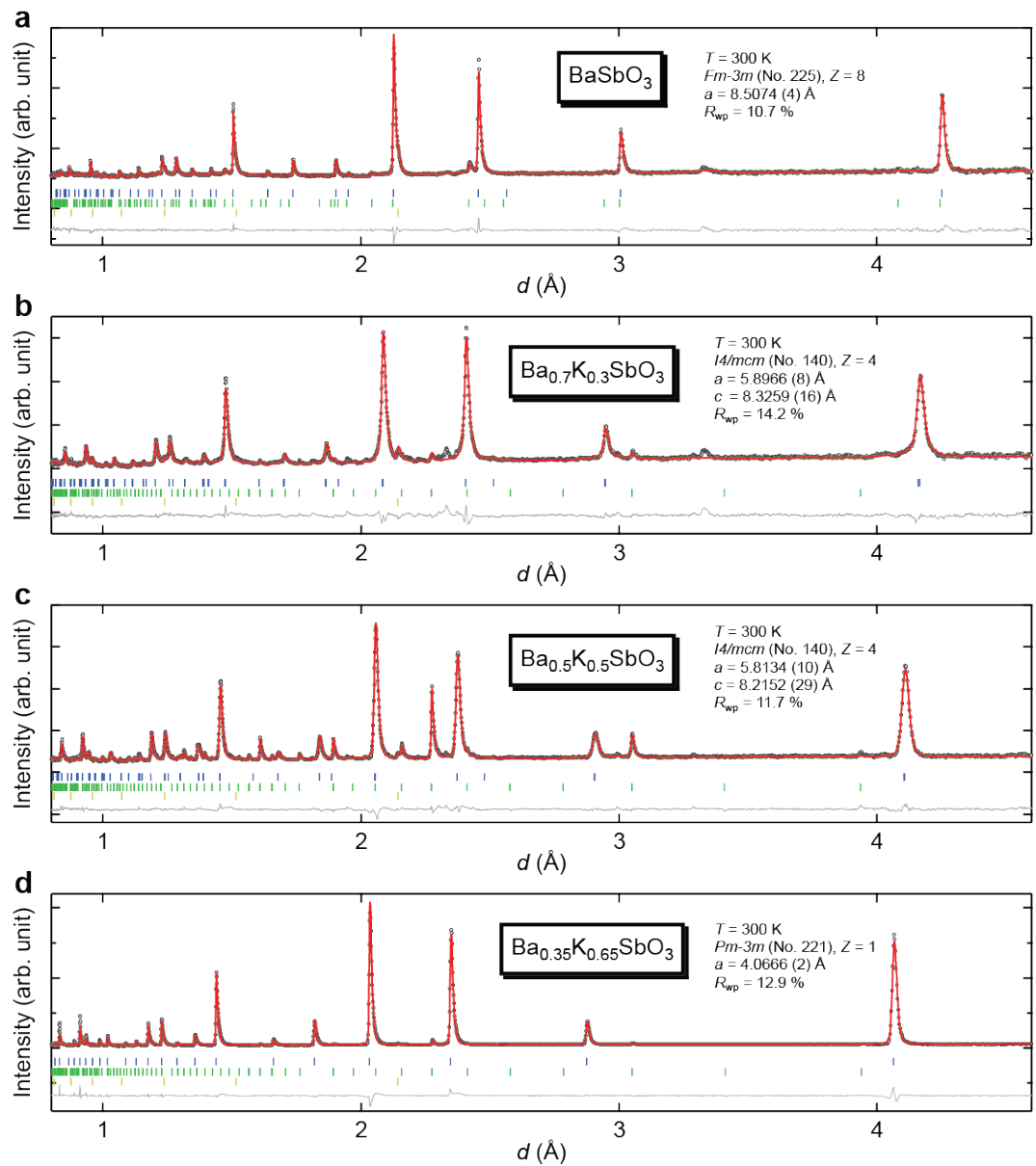

**Extended Data Figure 3 | Rietveld refinement profiles of diffraction patterns of BKSO samples based on powder neutron diffraction.** Neutron diffraction patterns of **a**, BaSbO$_{3-\delta}$, **b**, Ba$_{0.7}$K$_{0.3}$SbO$_3$, **c**, Ba$_{0.5}$K$_{0.5}$SbO$_3$, **d**, Ba$_{0.35}$K$_{0.65}$SbO$_3$ were collected at the WISH diffractometer, ISIS facility. Black dots are experimental data, red lines are the simulated intensity, and blue, green, and yellow ticks are the *hkl* indices of BaSbO$_{3-\delta}$, impurity phases (BaSbO$_{2.5}$ in BaSbO$_{3-\delta}$,

and $KSbO_3$ in the other three samples) and a vanadium sample can, respectively. Grey lines are the difference between the experimental data and the simulation.

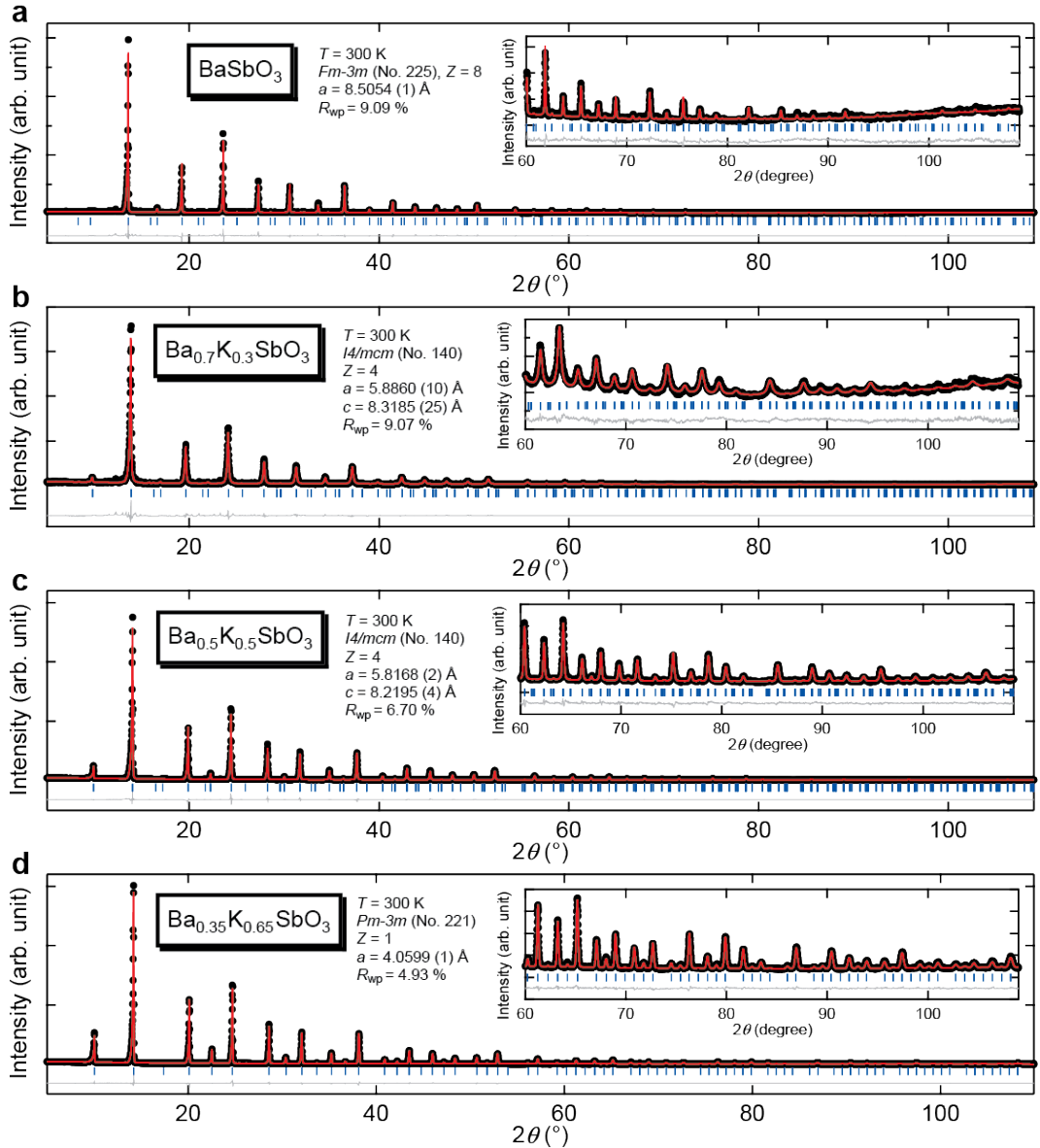

**Extended Data Figure 4 | Rietveld refinement profiles of diffraction patterns of BKSO samples based on powder X-ray diffraction.** X-ray diffraction patterns of of **a**, $BaSbO_{3-\delta}$, **b**, $Ba_{0.7}K_{0.3}SbO_3$, **c**, $Ba_{0.5}K_{0.5}SbO_3$, **d**, $Ba_{0.35}K_{0.65}SbO_3$ were collected in the Debye-Scherrer geometry with a Mo K$\alpha_1$ radiation at room temperature. Same notations as in Extended Data Figure 2 were used for plotting data.

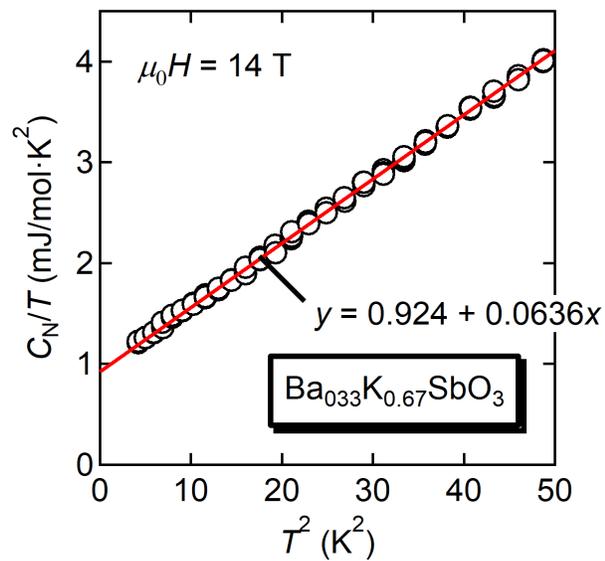

**Extended Data Figure 5 | Normal-state specific heat of the optimally doped BKSO.** Normal-state specific heat $C_N$ of Ba$_{0.33}$K$_{0.67}$SbO$_3$ was measured at the magnetic fields of 14 T, which is far larger than the upper critical field (~1 T) of the compound. A fit to the data with $C_N(T) = \gamma T + \beta T^3$ yields $\gamma = 0.924$ mJ·mol$^{-1}$K$^{-2}$ and $\beta = 0.0636$ mJ·mol$^{-1}$K$^{-4}$. From the value of $\beta$, the Debye temperature of 535 K was estimated.

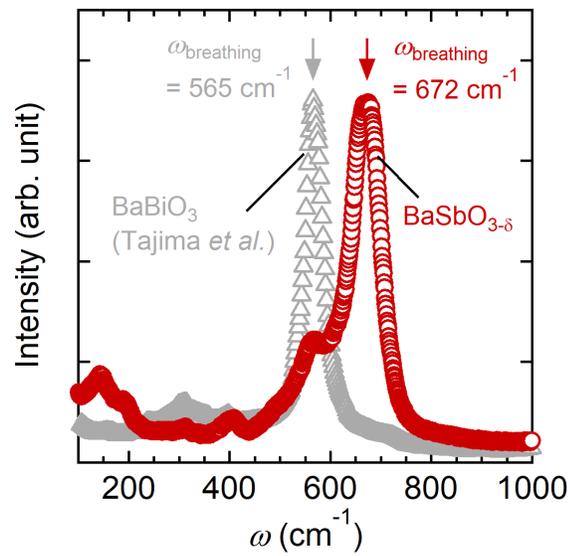

**Extended Data Figure 6 | Increased frequency of the breathing-mode phonon in the undoped antimonate as compared to the bismuthate.** The breathing-mode phonon of $BaSbO_{3-\delta}$ (red open circles) measured via Raman scattering shows 19 % increase of its frequency than that of $BaBiO_3$[35] (grey open triangles), due to the strong metal-oxygen covalency of the antimonates.